\documentclass[a4paper,11pt]{article}
\pdfoutput=1

\usepackage{jheppub} 
\usepackage{graphicx}
\usepackage{amsmath,amssymb,amsthm,amsfonts}
\usepackage{bbold}
\usepackage{float}
\usepackage{slashed}
\usepackage{dcolumn}
\usepackage{bm}
\usepackage{color}
\usepackage{multirow}
\usepackage{fontawesome}

\allowdisplaybreaks

\newcommand{\lsim}{{\;\raise0.3ex\hbox{$<$\kern-0.75em\raise-1.1ex\hbox{$\sim$}}\;}}
\newcommand{\gsim}{{\;\raise0.3ex\hbox{$>$\kern-0.75em\raise-1.1ex\hbox{$\sim$}}\;}}
\def\bea{\begin{eqnarray}}
\def\eea{\end{eqnarray}}
\def\bec{\begin{center}}
\def\ec{\end{center}}

\def\beq{\begin{equation}}
\def\eeq{\end{equation}}

\def\bea{\begin{eqnarray}}
\def\eea{\end{eqnarray}}
\def\beq#1\eeq{\begin{align}#1\end{align}}
\def\beqnn#1\eeq{\begin{align*}#1\end{align*}}
\def\ba{\begin{array}}
\def\ea{\end{array}}
\def\bc{\begin{center}}
\def\ec{\end{center}}

\def\gev{\rm GeV}
\def\tev{\rm TeV}

\usepackage{breqn} 

\usepackage{cases}

\newcommand{\be}{\begin{equation}\begin{aligned}}
\newcommand{\ee}{\end{aligned}\end{equation}}

\newcommand{\beqa}{\begin{eqnarray}}
\newcommand{\eeqa}{\end{eqnarray}}

\renewcommand{\eqref}[1]{Eq.~(\ref{#1})}

\newcommand{\eg}{{\em e.g.}}

\newcommand{\TODO}[1]{\textcolor{green}{TODO}}
\RequirePackage[normalem]{ulem}

\DeclareUnicodeCharacter{2212}{\textendash}

\makeatletter
\def\l@subsubsection#1#2{}
\makeatother

\preprint{CTPU-PTC-25-27}

\title{Dark axion portal at $Z$ boson factories}

\author{Krzysztof Jod\l{}owski}

\affiliation{Particle Theory and Cosmology Group, Center for Theoretical Physics of the Universe, \\ Institute for Basic Science (IBS), Daejeon 34126, Korea}

\emailAdd{k.jodlowski@ibs.re.kr}

\abstract{
    Dark axion portal connects an axion-like-particle (ALP), which can be the QCD axion, with the SM by the coupling to a photon and a dark photon, leading to a rich and distinct phenomenology related to dark matter, astrophysics, and cosmology. We note that due to the gauge invariance, the $Z$ boson-dark photon-ALP coupling should also be generated with sizable strength, and we study its phenomenological consequences at $Z$ boson factories: LEP, FCC-ee, and forward physics detectors at the LHC and FPF@FCC - FASER and MATHUSLA. Due to the large number of $Z$ bosons produced, DAP particles can be efficiently produced and detected by semi-visible displaced decays of the heavier DS species to $\gamma$+inv. or $f^+ f^-$+inv., and via missing energy signature with zero or one photon. Because of the complementarity of the two approaches, we find great prospects in exploring both short and long-lived lifetime regimes of DAP, especially when the heavier dark species has mass above $0.1\,\gev$.
}

\vspace{4cm}

\begin{document} 
\maketitle
\flushbottom

\section{Introduction}
\label{sec:introduction}
Dark axion portal (DAP) is an interesting example of a non-minimal dark sector (DS) model that can solve several pressing problems of cosmology and particle physics such as dark matter (DM), the strong-CP problem~\cite{Kaneta:2016wvf,Kaneta:2017wfh,Broadberry:2024yjw} or be related to the cosmological relaxation solution of the hierarchy problem \cite{Graham:2015cka,Choi:2016kke,Domcke:2021yuz}.
Since the DAP species - an axion $a$\footnote{Called further as an axion-like particle (ALP), since we also consider scenarios with a heavy ALP and a light DP, in which case $a$ generically is not expected to be the QCD axion.} and a dark photon (DP) $\gamma^\prime$ - are only feebly coupled to the SM, experimental probes of DAP require multifaceted approach.

In fact, it has been extensively studied recently in various physical settings, for example using probes from astrophysics~\cite{Hook:2021ous,Kalashev:2018bra}, \eg, supernovae or blazars, early Universe cosmology~\cite{Hook:2023smg,Hong:2023fcy}, \eg, CMB and BBN, or at terrestrial intensity frontier experiments such as beam dumps, neutrino experiments, B-meson factories, and forward physics detectors at the LHC~\cite{deNiverville:2018hrc,deNiverville:2019xsx,Deniverville:2020rbv,Jodlowski:2023yne}.
In this work, we point out that whenever the $g_{a\gamma \gamma^\prime}$ coupling is dominant, there is also a sizable $g_{a Z\gamma^\prime}$ coupling originating from the full SM gauge invariance.
Moreover, in the simplest scenario, this coupling is directly proportional to $g_{a\gamma \gamma^\prime}$, which is a different case than the $g_{a\gamma \gamma}$ and $g_{a\gamma Z}$ couplings of ordinary ALP, which do not need to be related to each other.

The interaction Lagrangian of the DAP is~\cite{Kaneta:2016wvf,Ejlli:2016asd}
\be
    \mathcal{L} &\supset \frac{g_{a\gamma \gamma^\prime}}{2}\, a\, F_{\mu\nu}  \tilde{F}_D^{\mu\nu}\,,
\label{eq:Lagrangian_DAP}
\ee
where $g_{a\gamma \gamma^\prime}$ is a coupling of mass-dimension $-1$, $a$ denotes an ALP, while the electromagnetic (EM) and the $U(1)_D$ tensors are given by $F^{\mu\nu}$ and $F_D^{\mu\nu}$, respectively. 
Moreover, $\tilde{H}^{\mu\nu}=\epsilon^{\mu\nu\rho\sigma}\, H_{\rho\sigma}/2$ is the field strength tensor dual to $H^{\mu\nu}$, where $H$ describes any gauge boson.

On the other hand, imposing the SM gauge invariance $SU(3) \times SU(2) \times U(1)$ means that Eq. \ref{eq:Lagrangian_DAP} is promoted to
\be
    \mathcal{L} &\supset \frac{g_{a\gamma \gamma^\prime}}{2}\, a\, F_{\mu\nu}  \tilde{F}_D^{\mu\nu} + \frac{g_{aZ \gamma^\prime}}{2}\, a\, Z_{\mu\nu}  \tilde{F}_D^{\mu\nu} \,, \,\,\mathrm{where} \,\,\, g_{aZ \gamma^\prime} = -\tan\theta_W\, g_{a\gamma \gamma^\prime} \,,
\label{eq:Lagrangian_DAP_EW}
\ee
and $\theta_W$ denotes the Weinberg angle.

Indeed, before the electroweak (EW) spontaneous symmetry breaking (SSB), the gauge-invariant Lagrangian for a light ALP with shift symmetry is~\cite{Bauer:2017ris,Bauer:2018uxu}
\be
    \mathcal{L} &\supset g_s^2 \, C_{GG} \, a\, G^A_{\mu\nu} \,\tilde{G}_A^{\mu\nu} + g^2\, C_{WW}\, a\, W^A_{\mu\nu} \,\tilde{W}_A^{\mu\nu} + g^{\prime \, 2}\, C_{BB}\, a\, B_{\mu\nu} \,\tilde{B}^{\mu\nu} + g_D^{\prime \, 2}\, C_{B B_D}\, a\, B_{\mu\nu} \,\tilde{B}_D^{\mu\nu} \,,
\label{eq:Lagrangian_before_SSB}
\ee
where $G$, $W$, $B$ describe the SM gauge bosons - gluon, weak gauge boson, and hypercharge boson with couplings $g_s$, $g$, $g'$, respectively, while $B_D$ describes the $U(1)_D$ gauge boson before the DS SSB.
After the EW SSB, $B_{\mu\nu} = \cos\theta_W \,F_{\mu\nu} - \sin\theta_W \, Z_{\mu\nu}$, and the Lagrangian given by Eq. \ref{eq:Lagrangian_before_SSB} reduces to Eq. \ref{eq:Lagrangian_DAP_EW} (the gluon coupling is not affected).

We note that while the ALP-dark photon couplings to photon/$Z$ boson have a fixed ratio - given by Eq. \ref{eq:Lagrangian_DAP_EW} - the ALP-photon couplings to photon/$Z$ boson are completely model dependent. 
In fact, they depend on the two parameters $C_{WW}$ and $C_{BB}$ \cite{Bauer:2018uxu} as follows:
\be
\label{eq:eeee}
    g_{aZ \gamma} = \frac{\cos\theta_W^2 C_{WW} - \sin\theta_W^2 C_{BB}}{\sin\theta_W \cos\theta_W (C_{WW} + C_{BB})} \, g_{a\gamma \gamma} \,.
\ee
We see that this relation reduces to Eq. \ref{eq:Lagrangian_DAP_EW} when $C_{WW}=0$. 

On the other hand, since the dark photon is abelian, it can only mix with hypercharge, and thus $|g_{a Z \gamma^\prime}/g_{a\gamma \gamma^\prime}|$ is fixed to $\tan\theta_W$. 
The leading correction to this relation comes from dimension-7 operator involving $W^a_\mu$ and two Higgs fields --  $(\partial_\mu a)(H^\dagger i\overleftrightarrow D_\nu H)\tilde F_D^{\mu\nu}$ is a shift-symmetric example.
We note that, by definition, DAP is characterized by negligible kinetic mixing $\mathcal{L} \supset \epsilon/2 F_{\mu\nu}  F_D^{\mu\nu}$, $\epsilon \simeq 0$.
Therefore, the mixing of the dark photon with $Z$ boson, which is proportional to $\epsilon$ - see, \eg,~\cite{Lee:2016ief} - is also negligible.

At momentum transfers and DS mass splittings well below $m_Z$, virtual-$Z$ effects are suppressed, and photon-mediated decays provide the leading approximation. 
The dominant collider consequence of $g_{aZ\gamma^{\prime}}$ in this regime is the additional production channel $Z\to a\gamma^{\prime}$. For larger mass splittings, which in this study are relevant only to LEP-2, the on-shell decays $\gamma^{\prime}\to Za$ or $a\to Z\gamma^{\prime}$ open.

Motivated by this, we investigate prospects of DAP at $Z$ boson factories, where one can employ techniques developed in studies devoted to, \eg, heavy neutral lepton (HNL), where analogous production mechanisms and decay channels take place.
In particular, the $e^+ e^-$ colliders running at the $Z$ mass center of mass energy, were shown, see, \eg, \cite{Zhang:2023nxy,Ovchynnikov:2023wgg,Drewes:2022rsk}, to provide significant bounds on such BSM scenario due to their large luminosities, lack of pileup, and controllable background.

This paper is organized as follows. In section \ref{sec:searches}, we introduce the DAP signatures that are relevant to the $Z$ boson factories and forward physics detectors at the LHC or FPF@FCC. We also describe our simulation setup and validate it for the HNL dipole portal. In section \ref{sec:results}, we present and discuss our results. We show that the LEP limits on DAP originating from the $g_{a Z\gamma^\prime}$ coupling put significant constraints, which are comparable or stronger than future Belle2 limits. We also present projected sensitivity of FCC-ee (further denoted as FCC), FASER2 and MATHUSLA, which will strongly improve the coverage of parameter space of large and small masses, respectively. Finally, we summarize our study in section \ref{sec:conclusions}.

\section{Signatures}
\label{sec:searches}
Since the DAP species, which can be produced in the $Z\to a+\gamma^\prime$ decays, are feebly coupled to the SM, they act as long-lived particles (LLP).
Following their production, we study their displaced decays into charged SM fermions and decays into a photon and an invisible state. 
We assume that interactions other than the mixed portal are negligible, and that the lighter dark state is effectively stable and invisible on detector scales. 

\paragraph{Displaced decays - decay widths}
The leading decay widths for the DS particles in the DAP are \cite{Kaneta:2016wvf}
\begin{equation}
    \Gamma_{\gamma^\prime \to \gamma a} = \frac{g_{a\gamma \gamma^\prime}^2}{96 \pi} m_{\gamma^{\prime}}^3 \left(1-\frac{m^2_a}{m^2_{ \gamma^\prime}}\right)^3, \,\,\,\, \Gamma_{a\to \gamma  \gamma^\prime} = \frac{g_{a\gamma \gamma^\prime}^2}{32 \pi} m_a^3 \left(1-\frac{m^2_{ \gamma^\prime}}{m^2_a}\right)^3.
\label{eq:Gamma_two_body}
\end{equation}

For displaced-vertex searches at lepton colliders, we consider three-body decays into charged SM fermions. Single-photon decays can also be probed using non-pointing photons~\cite{DELPHI:2003dlq}, but a dedicated displaced-photon analysis is beyond the scope of this work.
For final states involving EM charged leptons or quarks, the relevant decay widths read \cite{Jodlowski:2023yne}:
\begin{equation}
    \begin{split}
    \Gamma_{\gamma^\prime \to a f\bar{f}} \simeq Q_f^2 N_f\,\frac{\alpha_{\mathrm{EM}} g_{a\gamma  \gamma^\prime}^2}{576 \pi ^2 m_{\gamma^{\prime}}^3} \biggl(& m_{\gamma^{\prime}} \sqrt{m_{\gamma^{\prime}}^2-4 m_f^2} \left(26 m_{\gamma^{\prime}}^2 m_f^2-7 m_{\gamma^{\prime}}^4+8 m_f^4\right) \\
    &  -4 m_{\gamma^{\prime}}^6 \log \left(\frac{2 m_f}{\sqrt{m_{\gamma^{\prime}}^2-4 m_f^2}+m_{\gamma^{\prime}}}\right) +32 m_f^6 \coth^{-1}\left(\frac{m_{\gamma^{\prime}}}{\sqrt{m_{\gamma^{\prime}}^2-4 m_f^2}}\right) \\
    & + 12 m_{\gamma^{\prime}}^2 m_f^4 \log \left(\frac{16 m_f^4 \left(m_{\gamma^{\prime}}-\sqrt{m_{\gamma^{\prime}}^2-4 m_f^2}\right)}{\left(\sqrt{m_{\gamma^{\prime}}^2-4 m_f^2}+m_{\gamma^{\prime}}\right)^5}\right)\biggr), \\
    \Gamma_{a\to \gamma^\prime f\bar{f}} \simeq Q_f^2 N_f\,\frac{\alpha_{\mathrm{EM}} g_{a\gamma  \gamma^\prime}^2}{192 \pi ^2 m_a^3} \biggl(&32 m_f^6 \coth^{-1}\left(\frac{m_a}{\sqrt{m_a^2-4 m_f^2}}\right) \\
    &+ m_a\sqrt{m_a^2-4 m_f^2} \left(26 m_a^2 m_f^2-7 m_a^4+8 m_f^4\right) \\
    &-4 m_a^6 \log \left(\frac{2 m_f}{\sqrt{m_a^2-4 m_f^2}+m_a}\right) \\
    & +12 m_a^2 m_f^4 \log \left(\frac{16 m_f^4 \left(m_a-\sqrt{m_a^2-4 m_f^2}\right)}{\left(\sqrt{m_a^2-4 m_f^2}+m_a\right)^5}\right) \biggr)
\,.
\label{eq:A_affbar_a_Affbar}
\end{split}
\end{equation}
These tree-level photon-mediated widths neglect the mass of the lighter DS particle.\footnote{Since the origin of the dark photon and axion masses are different, we consider scenarios where either $m_{\gamma^{\prime}}\gg m_a$ or $m_a\gg m_{\gamma^{\prime}}$.} 
Here $Q_f$ denotes the electric charge in units of $e$, and the color factor is $N_f=1$ for leptons and $N_f=3$ for quarks. The leptonic expressions apply directly, while the quark expressions are partonic approximations. 
The plotted decay widths correspond to virtual-photon channels only. 
For charged final states and invariant masses $q^2\ll m_Z^2$, the pure virtual-$Z$ contribution relative to photon exchange is suppressed by $(q^2/m_Z^2)^2$, whereas their interference begins at order $q^2/m_Z^2$. 
Above the two-body thresholds, the additional widths are
\begin{align*}
    \Gamma_{\gamma^{\prime}\to Za} &= \frac{g_{aZ\gamma^{\prime}}^2}{96\pi m_{\gamma^{\prime}}^3} \lambda^{3/2}(m_{\gamma^{\prime}}^2,m_a^2,m_Z^2)
\,,
\\
    \Gamma_{a\to Z\gamma^{\prime}} &= \frac{g_{aZ\gamma^{\prime}}^2}{32\pi m_a^3} \lambda^{3/2}(m_a^2,m_{\gamma^{\prime}}^2,m_Z^2)
\,,
\end{align*}
where $\lambda(x,y,z)=x^2+y^2+z^2-2xy-2xz-2yz$. 

\begin{figure}[tb]
    \centering
    \includegraphics[scale=0.36]{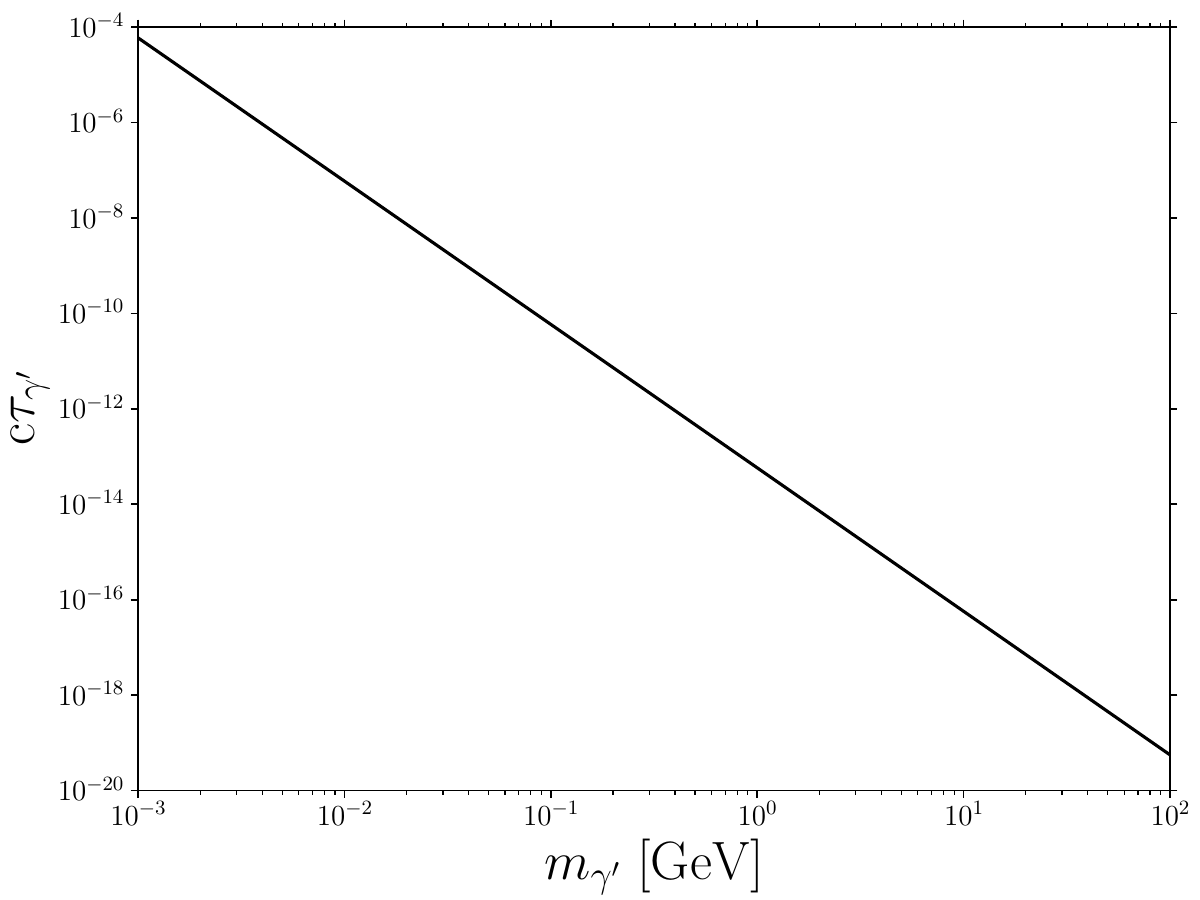}\vspace{0.25cm}
    \includegraphics[scale=0.36]{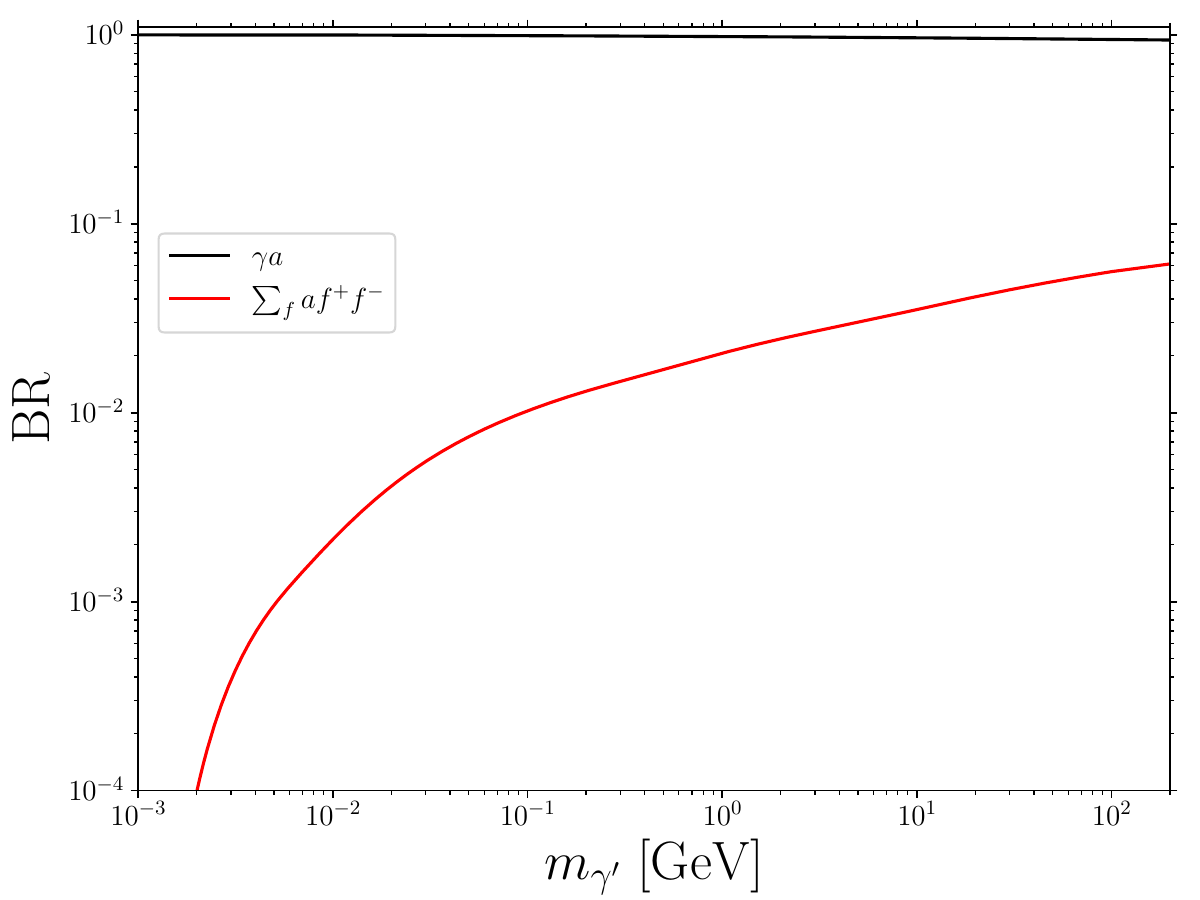}
    \includegraphics[scale=0.36]{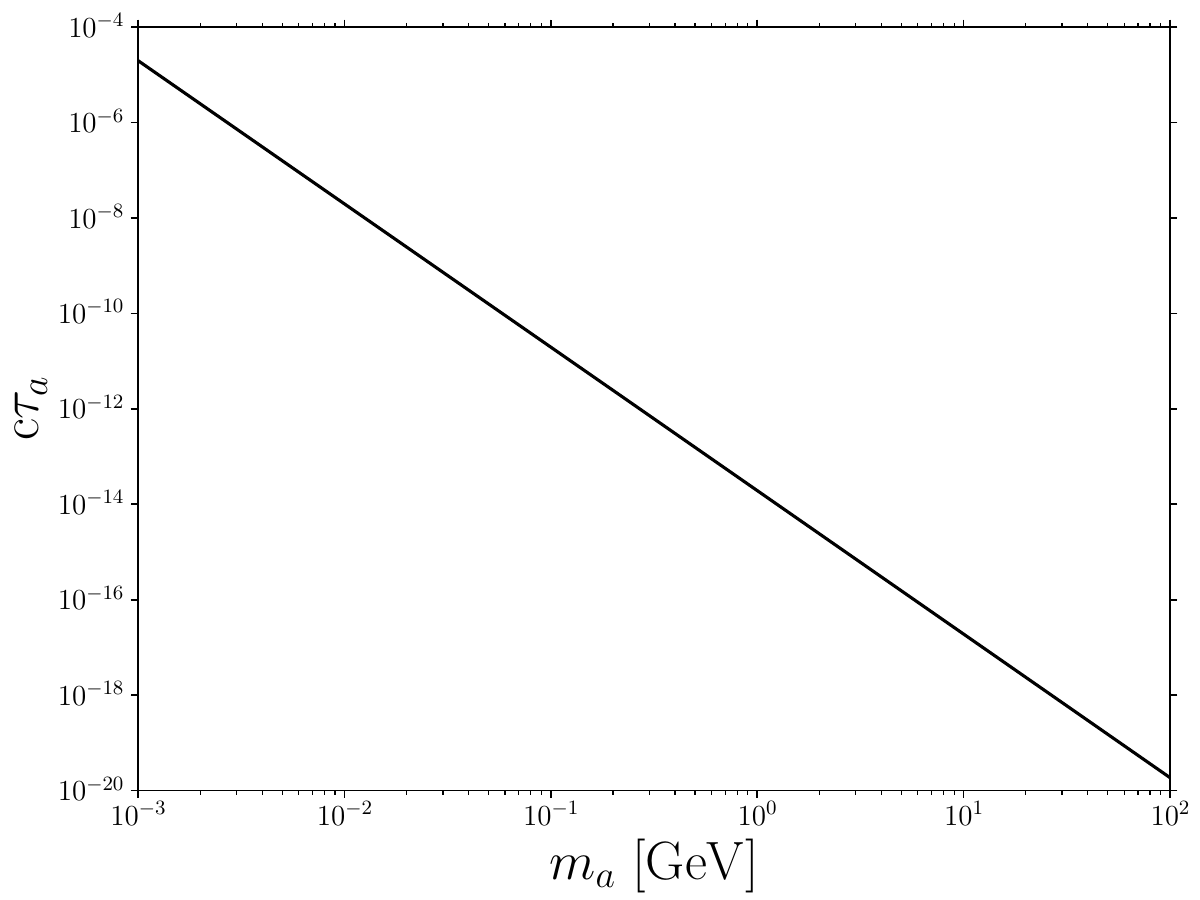} \vspace{0.25cm}
    \includegraphics[scale=0.36]{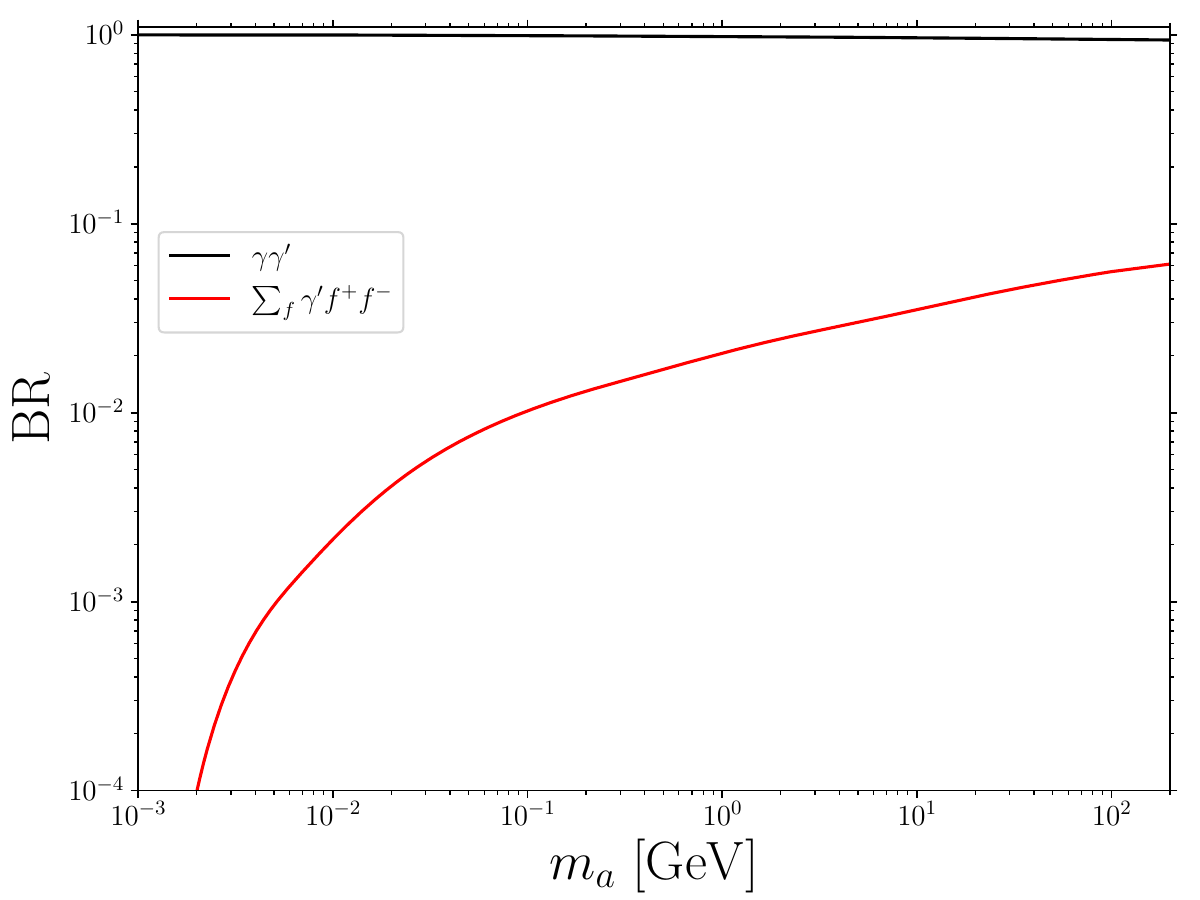}
    \caption{
            Photon-mediated lifetimes, $c\tau\,[\mathrm{m}]$ (left), and branching fractions (right), as functions of the LLP mass for the two DAP mass hierarchies.
            The top panels correspond to $m_{\gamma'} \gg m_a$, while bottom panels illustrate $m_a \gg m_{\gamma'}$.
            In the left panels, we fixed $g_{a\gamma\gamma'}=1/$GeV, while the general case is obtained by rescaling by $1/g_{a\gamma\gamma'}^2$.
        }
        \label{fig:lifetime_BR}
    \end{figure}
In Fig.~\ref{fig:lifetime_BR}, we present the dependence of lifetime and branching ratio on the mass of the considered LLP on the left and right panels, respectively.
The two-body decays dominate over the whole mass range, however, although the three-body decays are suppressed by $\sim\alpha_{\mathrm{EM}}/(3\pi)$, with logarithmic enhancement, this is partially compensated by the number of degrees of freedom of EM charged fermions. 
As a result, the semi-visible decays into charged fermions contribute at over percent level for $m\gtrsim 1\gev$.
The three-body channel is relevant for searches that rely on charged-particle vertices, in particular, for MATHUSLA, which is planned to contain trackers but no calorimeter, and the displaced decays searches at the FCC.

\paragraph{Production}
For $m_a+m_{\gamma^{\prime}}<m_Z$, DAP particles can be produced by on-shell $Z$ decays with the following width:
\begin{equation}
    \Gamma_{Z \to \gamma^\prime a} = \frac{g_{a Z \gamma^\prime}^2}{96\pi m_Z^3} \left[(m_a-m_{\gamma^{\prime}}-m_Z) (m_a+m_{\gamma^{\prime}}-m_Z) (m_a-m_{\gamma^{\prime}}+m_Z) (m_a+m_{\gamma^{\prime}}+m_Z) \right]^{3/2}  \,.
\label{eq:Z_ALPDP}
\end{equation}

Lepton colliders run not only at the $Z$ boson peak, but also at higher energies. There, the dominant production mode comes from $e^+ e^-$ annihilation into $a$ and $\gamma^\prime$. 
For real couplings, unpolarized beams, and negligible electron mass, coherent photon and $Z$ exchange give the following cross section for $\sqrt{s}>m_a+m_{\gamma^{\prime}}$:
\begin{equation}
\begin{split}
    \sigma_{e^{+}e^{-}\to\gamma^{*}/Z^{*}\to a\gamma^{\prime}} ={}& \frac{\alpha_{\mathrm{EM}}} {192\cos^{2}\theta_W\sin^{2}\theta_W\, s^{3}\bigl((s-m_Z^{2})^{2}+m_Z^{2}\Gamma_Z^{2}\bigr)}
    \\
    &\times \left(m_a^{4}-2m_a^{2}(m_{\gamma^{\prime}}^{2}+s) +(m_{\gamma^{\prime}}^{2}-s)^{2}\right)^{3/2}
    \\
    &\times\Bigl( 8\cos^{2}\theta_W\sin^{2}\theta_W\, g_{a\gamma\gamma^{\prime}}^{2} \bigl((s-m_Z^{2})^{2}+m_Z^{2}\Gamma_Z^{2}\bigr)
    \\
    &\hspace{15mm} +g_{aZ\gamma^{\prime}}^{2}s^{2} (8\sin^{4}\theta_W-4\sin^{2}\theta_W+1)
    \\
    &\hspace{15mm} +4\sin\theta_W\cos\theta_W(1-4\sin^{2}\theta_W)
    \\
    &\hspace{25mm}\times g_{a\gamma\gamma^{\prime}}g_{aZ\gamma^{\prime}}\, s(s-m_Z^{2}) \Bigr)
\,.
\label{eq:ee_to_aDP}
\end{split}
\end{equation}
Setting $g_{aZ\gamma^{\prime}}=0$, recovers the photon-mediated result of Ref.~\cite{deNiverville:2018hrc} (Eq.~6 therein).
We also checked that the impact of the $g_{a Z \gamma^\prime}\neq 0$ coupling on the results shown in Fig. 4 therein is negligible because the B-factories run at $\sqrt{s}\simeq 10\,\gev \ll m_Z$.

\paragraph{Simulation setup and validation}
Let us note the following similarity\footnote{On the other hand, DAP is dimension-5 coupling with \textit{fixed} ratio of $g_{a\gamma\gamma'}/g_{a Z \gamma'}$, while HNL magnetic dipole portal before the electroweak spontaneous symmetry breaking requires Higgs doublet. Therefore, it is actually a dimension-6 coupling, where the ratio $d_{N\nu\gamma}/d_{N\nu Z}$ is a free variable.}
between the dark axion and HNL magnetic dipole portals, which are described by Eq. \ref{eq:Lagrangian_DAP} and $\mathcal{L}_{\text{dipole}} = d_\alpha \, \bar{N} \sigma^{\mu\nu} \nu_{\alpha L} \, F_{\mu\nu} + \text{h.c.}$, respectively. 
Both portals yield analogous photon-plus-missing-energy and displaced charged-fermion signatures, motivating the use of the search strategies in Refs.~\cite{Zhang:2023nxy,Ovchynnikov:2023wgg}. 
The different spins lead to different production and decay angular distributions. The HNL comparison below tests our implementation and detector treatment, while a dedicated validation of the DAP efficiencies, including production--decay spin correlations, is beyond the scope of this work. 
We employ the same parameters and cuts as described in these works; \eg, for the FCC, see Tab. 1 and 3 in~\cite{Ovchynnikov:2023wgg}, while for LEP - see Eq. 18, 20 and 21 in \cite{Zhang:2023nxy}.
We also use the following luminosities and the lengths of the detectors. FCC: $\mathcal{L}_{\mathrm{FCC}}=145\,$ab$^{-1}$, $l_{\mathrm{FCC}}=5.44\,m$, LEP: $\mathcal{L}_{\mathrm{LEP}}=0.2\,$fb$^{-1}$ and $l_{\mathrm{LEP}}=1\,m$.

\begin{figure}[tb]
    \centering
    \includegraphics[scale=0.4]{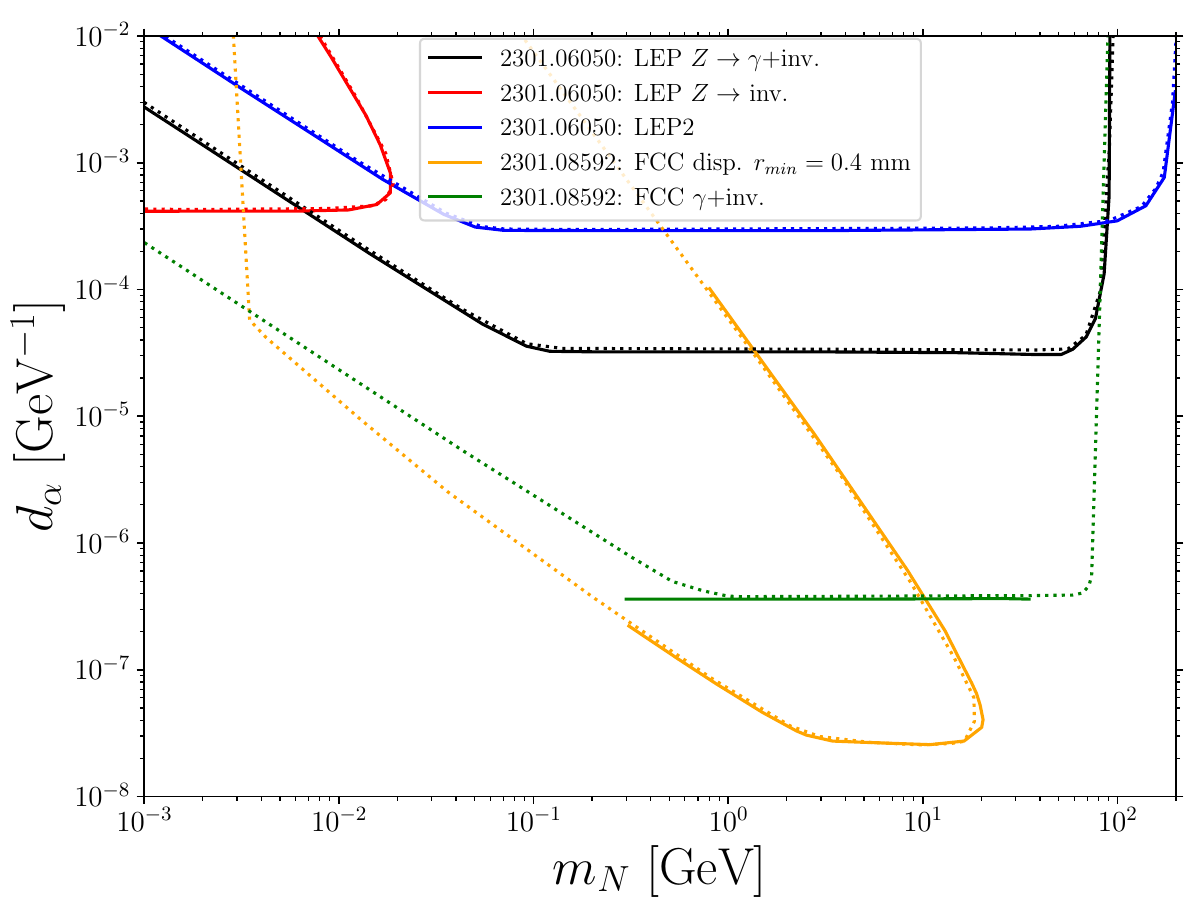}
    \caption{
        Sensitivity plot for HNL dipole portal obtained in \cite{Zhang:2023nxy,Ovchynnikov:2023wgg} (solid lines), which we used to validate our simulation setup.
        Whenever relevant, we recovered (dotted lines) each of the limits.
        The red, blue, and black solid lines show the LEP constraints recast from Ref.~\cite{Zhang:2023nxy}. The orange and green lines indicate the FCC-ee displaced-decay and monophoton projections of Ref.~\cite{Ovchynnikov:2023wgg}.
        }
    \label{fig:validation_1}
\end{figure}

In Fig. \ref{fig:validation_1}, we show the results of our simulation employed for the HNL dipole portal in order to  validate it. 
We recovered both the LEP bounds found in \cite{Zhang:2023nxy} (see Fig. 3 scenario I therein) and the projected sensitivity of FCC found in \cite{Ovchynnikov:2023wgg} (see Fig. 8 therein).
We note that the only discrepancy we found was at the low mass region, $m_{\mathrm{HNL}} \equiv m_{N} \lesssim 1\,\gev$, for the FCC $\gamma$+inv. search, which is denoted by green line. 
We checked that if one omits the factor describing probability of HNL decay into $\gamma$ and neutrino, see Eq. 18 in \cite{Zhang:2023nxy}, we recover the solid green line.

\section{Results}
\label{sec:results}
We consider two benchmarks of DAP: $m_{\gamma^{\prime}}\gg m_a$ and $m_a\gg m_{\gamma^{\prime}}$. At equal LLP mass and portal coupling, with a massless final dark state, the radiative and photon-mediated three-body widths of the DP are one third of the corresponding ALP widths. The photon-mediated DP lifetime is therefore three times longer.

\begin{figure}[tb]
    \centering
    \includegraphics[scale=0.36]{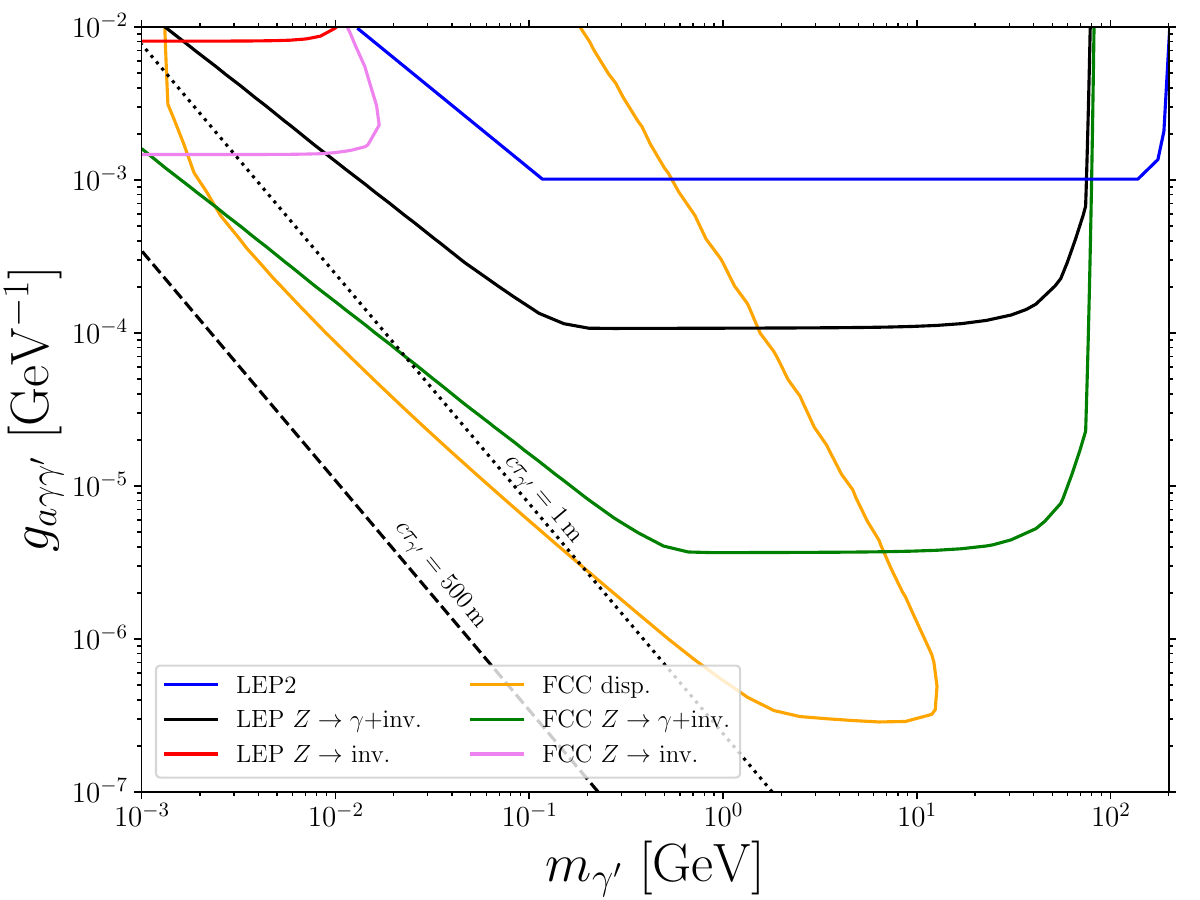}\vspace{0.25cm}
    \includegraphics[scale=0.36]{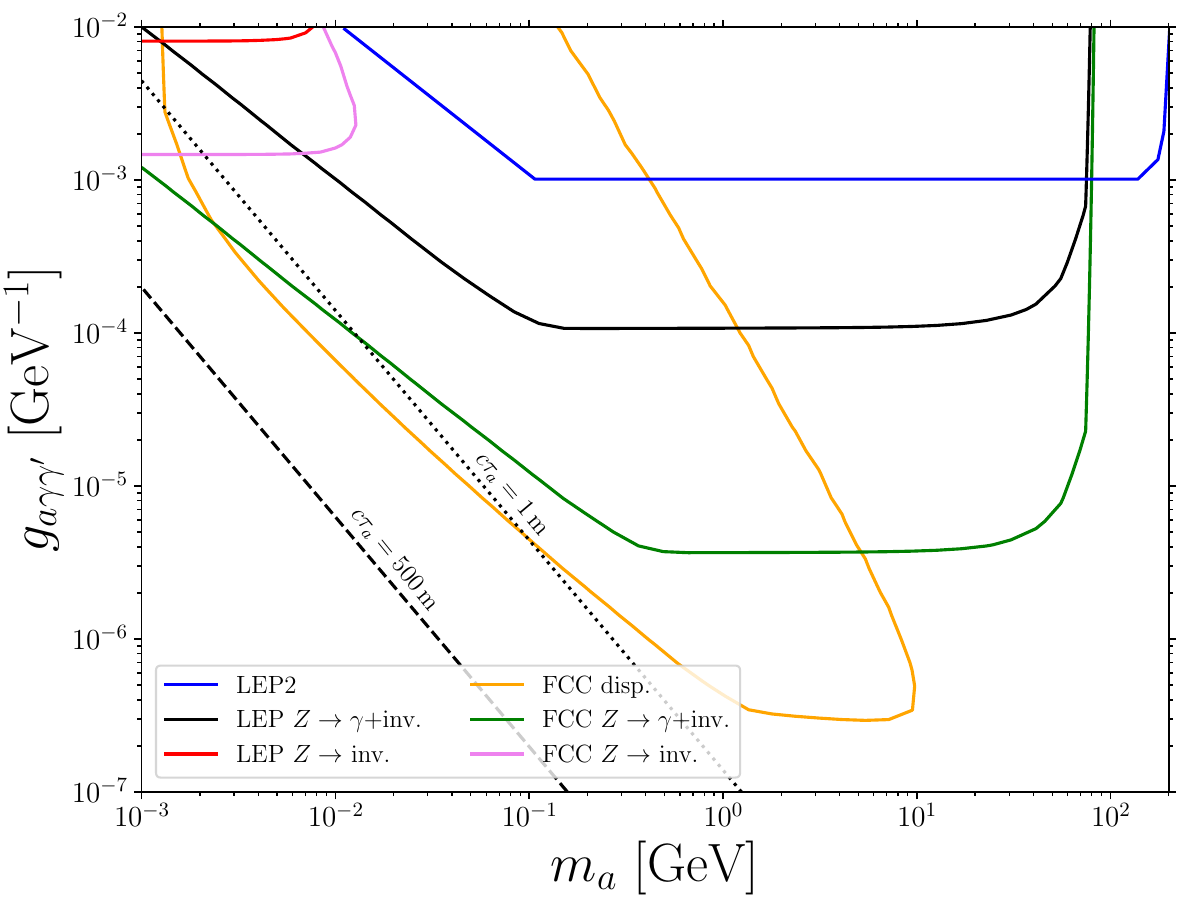}
    \caption{
        Lepton colliders sensitivity to DAP with massless ALP (left) and massless DP (right).
        LEP exclusion bounds are shown in red, blue and black, while future sensitivity of FCC is shown in pink, orange and green.
        The search strategies follow those used for HNL; comparison is shown in Fig.~\ref{fig:validation_1}. 
        The curves involving observed decay products are conditional on the DAP acceptance treatment discussed in the text, and the high-mass LEP2 contour uses photon-mediated decay approximation. 
        }
    \label{fig:result_1}
\end{figure}

In Fig.~\ref{fig:result_1}, we present the results of our simulation for lepton colliders searches for DAP. 
We also present contour lines for fixed values of LLP lifetime, $c\tau=500\,m$ (dashed) and $c\tau=1\,m$ (dotted). 
These proper-lifetime contours provide reference benchmarks. Sensitivity depends on the boosted decay length $\beta\gamma c\tau$ and on the observable event yield, while invisible searches remain applicable in the long-lived regime.
We find that LEP running at $\sqrt{s}=m_Z$ can put strong constraints by employing the $Z\to \gamma$+inv. signature, while LEP2 running at $\sqrt{s}=205.4\,\gev$, can extend the limits to higher masses.
On the other hand, lower values of the masses are covered by the $Z\to$ inv.

FCC will not only improve upon these LEP bounds due to its gain in luminosity, but actually it will set the most stringent bound via displaced decays signature, assuming background rejection and perfect signal efficiencies.
In fact, these would be the strongest terrestrial limits on DAP in the $m \gtrsim 1\,\gev$ mass region.

For a monophoton search with a large expected background and negligible systematic uncertainties, the limiting accepted signal cross section scales approximately as $\mathcal{L}^{-1/2}$ if the accepted background rate per unit luminosity is comparable at LEP and FCC-ee,\footnote{Cf. Eq.~2.12 in Ref.\cite{Ovchynnikov:2023wgg}.}
\be
    \frac{\sigma_{\mathrm{limit}}^{\mathrm{FCC}}} {\sigma_{\mathrm{limit}}^{\mathrm{LEP}}} \simeq \sqrt{\frac{\mathcal{L}_{\mathrm{LEP}}} {\mathcal{L}_{\mathrm{FCC}}}}
\,,
\ee
For the $Z\to\mathrm{inv.}$ signature, the projected sensitivity also depends on the experimental precision of the invisible-width measurement and on the LLP escape probability. A larger detector reduces this probability when the boosted decay length is comparable to the detector dimensions.

\begin{figure}[tb]
    \centering
    \includegraphics[scale=0.36]{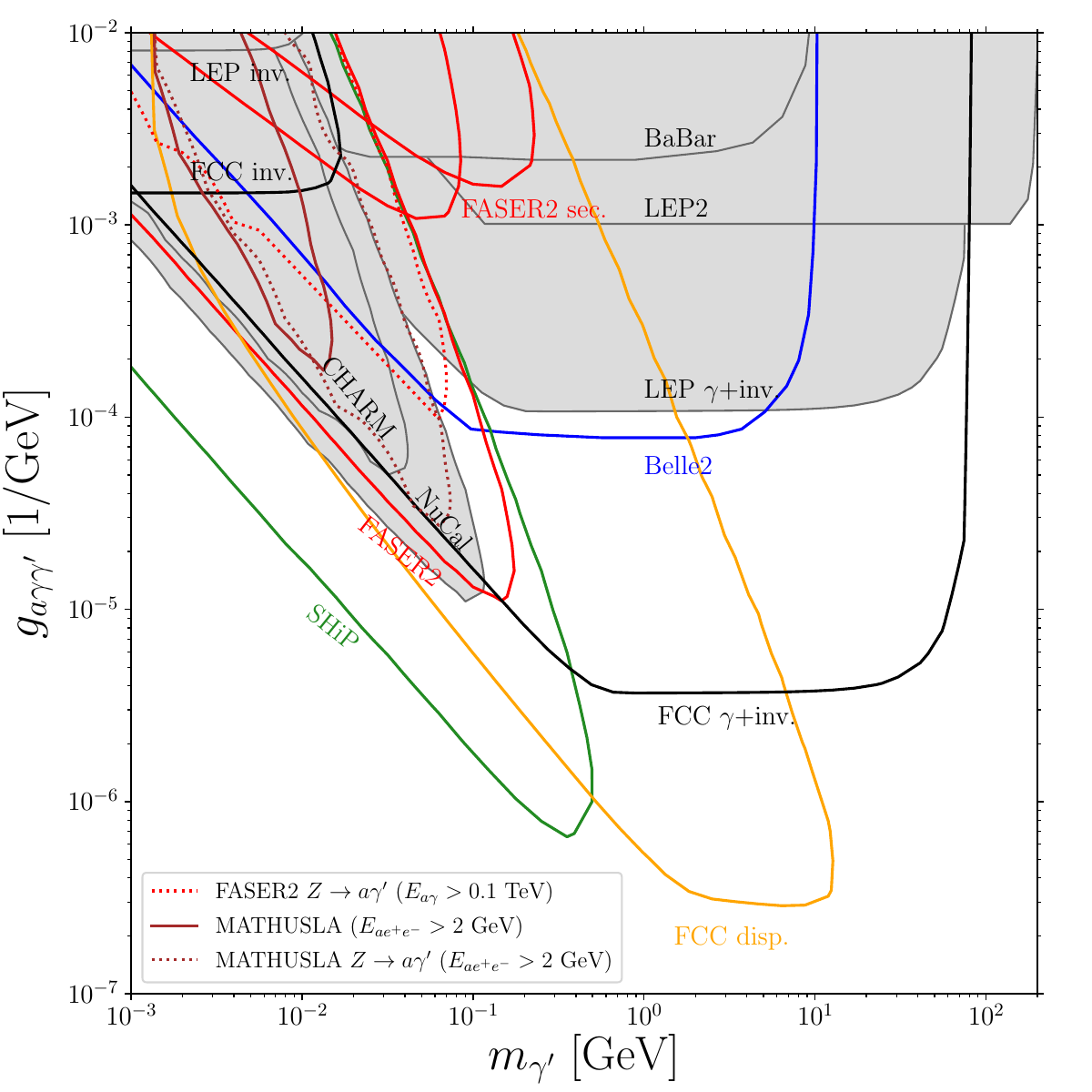}\vspace{0.35cm}
    \includegraphics[scale=0.36]{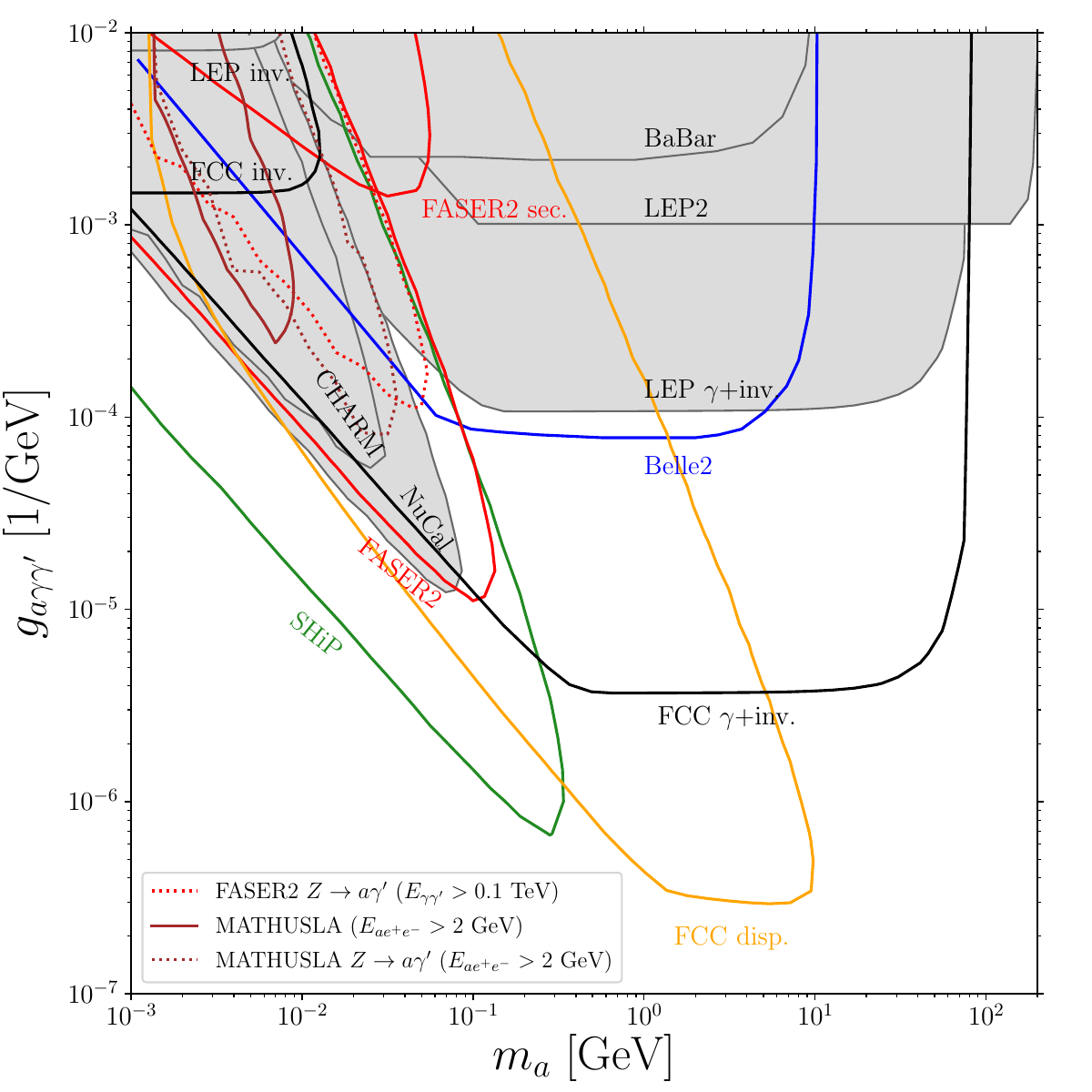}\\
    \includegraphics[scale=0.36]{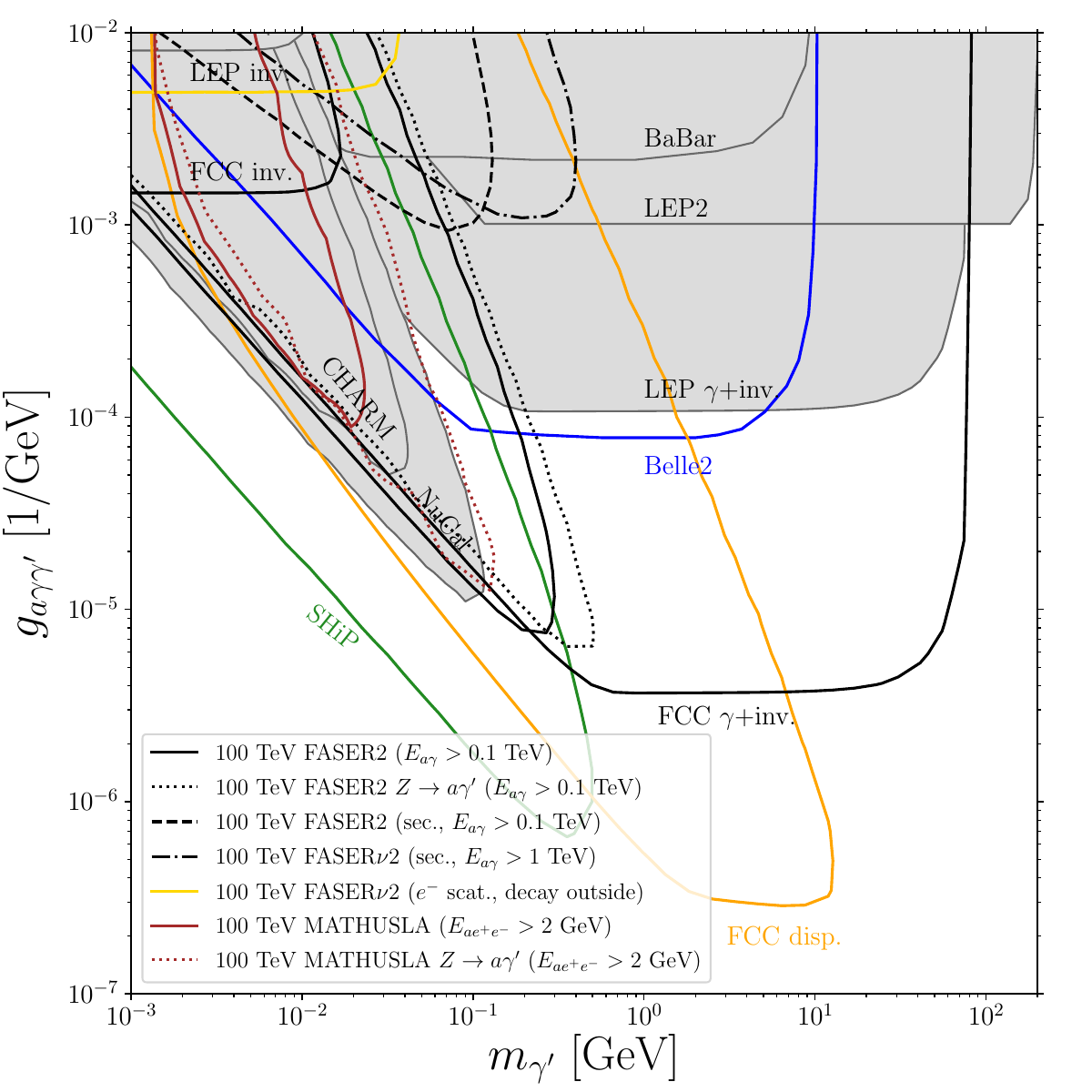}\vspace{0.35cm}
    \includegraphics[scale=0.36]{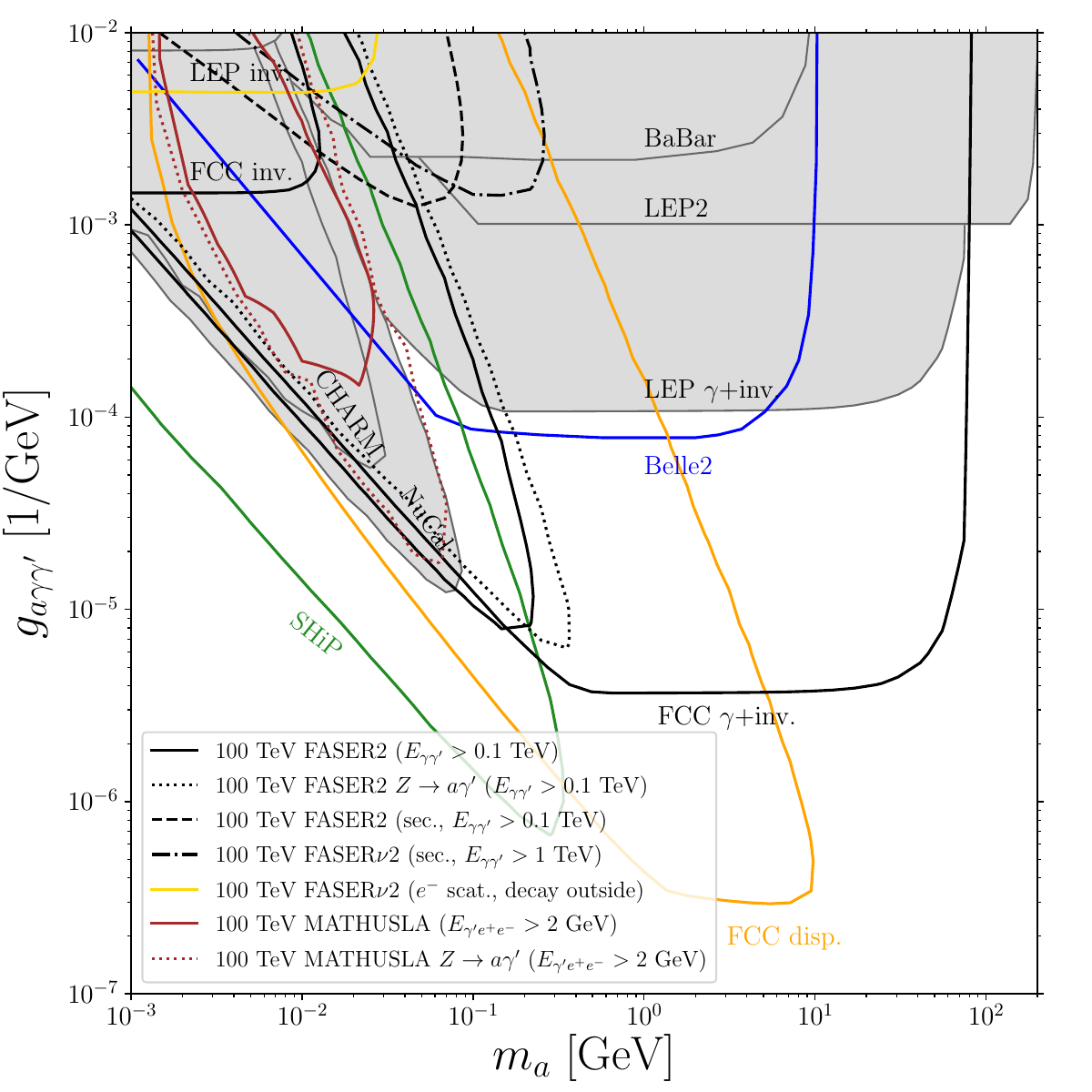}
    \caption{
        Combined lepton colliders, beam dumps, and LHC (top) or FPF@FCC (bottom) forward physics detectors sensitivity to DAP with massless ALP (left) and massless DP (right).
        The lepton-collider results are those shown in Fig.~\ref{fig:result_1}.
        The limits shown in legends of each plot are results of this work, while other limits were obtained in \cite{deNiverville:2018hrc,Jodlowski:2023yne}.
        Since at the FPF@FCC the $Z\to a \gamma^\prime$ production mode is dominant, we have shown resulting limits by dotted lines separately from those obtained by vector meson decays.
        For clarity, we have omitted astrophysical and cosmological bounds~\cite{Hook:2021ous,Kalashev:2018bra,Hook:2023smg,Hong:2023fcy} relevant for $m \lesssim 1\,\gev$ and small values of $g_{a\gamma \gamma^\prime}$.
        }
    \label{fig:result_2}
\end{figure}

In Fig.~\ref{fig:result_2}, we show our combined results, which include these already shown in Fig.~\ref{fig:result_1}, and bounds set by forward physics detectors at the LHC or at the proposed $100\,\tev$ FPF@FCC~\cite{MammenAbraham:2024gun} facility - FASER2 \cite{FASER:2018ceo,FASER:2018bac,FASER:2021ljd} and MATHUSLA\footnote{Unlike FASER, MATHUSLA will not be sensitive to LLP decays to photons, hence its sensitivity is suppressed by the three-body branching ratio, see Fig. \ref{fig:lifetime_BR}. On the other hand, its geometry allows to cover the large transverse momentum regime, complementary to the highly collimated LLPs covered by FASER.} \cite{Chou:2016lxi,Curtin:2018mvb}.
To obtain their projected limits, we implemented DAP with the $g_{a Z\gamma^\prime}$ coupling in modified $\tt FORESEE$ \cite{Kling:2021fwx} package.
The projected sensitivity curves correspond to three accepted signal events (approximate 95\% Poisson exclusion threshold) with the cuts indicated in Tab.~1 of Ref.~\cite{Jodlowski:2023yne}. 
Within the used photon-mediated partonic approximation, the summed three-body contribution reaches approximately $6$ percent for $m_{a/\gamma'}=200\,\mathrm{GeV}$. 
At the LHC, heavy vector meson decays - mostly $J/\psi$, see Fig. 1 in \cite{Jodlowski:2023yne} - are the leading production mode at forward detectors, while the $Z\to a+\gamma^\prime$ decays are subdominant.
On the other hand, at the $100\,\tev$ FCC collider the latter decays dominate - cf. solid and dotted lines in Fig.~\ref{fig:result_2}.

\section{Conclusions}
\label{sec:conclusions}
Dark axion portal is a well-motivated DS model that combines axion and vector portals in a non-trivial fashion, resulting in an interesting phenomenology. 
While its defining feature is the dominant ALP-photon-dark photon coupling, we have shown that in order to preserve the SM gauge invariance before the SSB, the coupling of the DS species to the $Z$ boson will also be generated with comparable strength, $g_{aZ\gamma^\prime}=-\tan\theta_W\,g_{a\gamma\gamma^\prime}$. 

We have investigated the resulting phenomenology, where the biggest impact on prospects of DAP searches utilizing $g_{a Z\gamma^\prime}$ coupling stems from the $Z$ boson factories. 
In particular, we considered lepton colliders running at the $Z$ mass center of mass energy such as LEP and FCC-ee. We adapted the methodology developed in~\cite{Zhang:2023nxy,Ovchynnikov:2023wgg,Drewes:2022rsk,Zhang:2022spf} to probe dipole portal HNL to the DAP. 
Under stated assumptions, we found that limits from LEP set the most stringent bounds in the $m \gtrsim 0.1\,\gev$ mass region, while FCC-ee will improve upon them by more than an order of magnitude not only by employing $Z$ boson decays into invisible states or a single photon with missing energy, but also by displaced decays signature.

Finally, we determined the yield and number of displaced decays of DAP species coming from $Z\to a+\gamma^\prime$ decays taking place at the LHC and the proposed $100\,\tev$ FPF@FCC facility at the FASER2 and MATHUSLA detectors.
For the $100\,\tev$ collider, the three-event projections illustrate the complementary reach of $Z$ and vector-meson production in the assumed FASER2 and MATHUSLA configurations.

In summary, we have shown that the $g_{a Z\gamma^\prime}$ coupling allows us to efficiently probe the DAP, especially when $m \gtrsim 0.1\,\gev$.
In particular, we found that LEP has put the strongest terrestrial limits covering several orders of magnitude in mass and coupling, while FCC will significantly improve them under optimistic assumptions regarding signal efficiencies and experimental backgrounds.

\section*{Acknowledgments}
This work was supported by IBS under the project code, IBS-R018-D1. 

\bibliography{bibliography}
\bibliographystyle{utphys}

\end{document}